# A Case Study of Nonresponse Bias Analysis in Educational Assessment Surveys

**Corresponding author:** Yajuan Si, Research Assistant Professor,
Survey Research Center, Institute for Social Research, University of Michigan, Ann Arbor
ISR 4014, 426 Thompson St, Ann Arbor, MI 48104
Email: yajuan@umich.edu; Phone: 734-7646935; Fax: 734-7648263

YAJUAN SI is a Research Assistant Professor in the Survey Research Center of the Institute for Social Research at the University of Michigan, ISR 4014, 426 Thompson St, Ann Arbor, MI 48104; Email: yajuan@umich.edu. Her research interests include Bayesian statistics, survey inference, missing data imputation, and data confidentiality protection.

RODERICK J.A. LITTLE is the Richard D. Remington Distinguished University Professor in the Department of Biostatistics, Research Professor in the Institute for Social Research, and Professor in the Department of Statistics at the University of Michigan, M4071 SPH II, 1415 Washington Heights, Ann Arbor, Michigan 48109; Email: rlittle@umich.edu. His research interests include incomplete data, sample surveys, Bayesian statistics, and applied statistics.

YA MO is an assistant professor of Curriculum, Instruction, and Foundational Studies at Boise State University and a research fellow at the National Institute of Statistical Sciences. 1750 K Street, NW, Suite 1100, Washington, DC 20006; Email: ymo@niss.org. Her research interests include quantitative methods, psychometric measures, survey statistics, and large-scale educational assessments.

NELL SEDRANSK is the Director of the DC Office of the National Institute of Statistical Sciences and Professor of Statistics at North Carolina State University. 1750 K Street, NW, Suite 1100, Washington, DC 20006; Email: NSedransk@niss.org. Her research interests include design of complex experiments, Bayesian inference, spatial statistics, and topological foundations for statistical theory.

# A Case Study of Nonresponse Bias Analysis in Educational Assessment Surveys

## Abstract

Nonresponse bias is a widely prevalent problem for data on education. We develop a ten-step exemplar to guide nonresponse bias analysis (NRBA) in cross-sectional studies and apply these steps to the Early Childhood Longitudinal Study, Kindergarten Class of 2010-11. A key step is the construction of indices of nonresponse bias based on proxy pattern-mixture models for survey variables of interest. A novel feature is to characterize the strength of evidence about nonresponse bias contained in these indices, based on the strength of the relationship between the characteristics in the nonresponse adjustment and the key survey variables. Our NRBA improves existing methods by incorporating both missing at random and missing not at random mechanisms, and all analyses can be done straightforwardly with standard statistical software.

1. Introduction

Surveys of educational assessment, such as the National Assessment of Educational Progress survey, the Program for International Student Assessment, and the Early Childhood Longitudinal Study (ECLS), have provided important data to inform policymakers and improve educational experiences. Such large-scale studies often implement complex probability sample designs to collect educational measurements on a sample representative of the target population. Survey variables are only collected for respondents to the study. However, statistical analyses of the collected data are subject to nonresponse bias, especially given rapidly declining response rates. A variety of indicators of potential bias have been proposed, generally functions of the response rate and the difference between respondents and nonrespondents on variables measured for both respondents and nonrespondents, such as auxiliary variables available in the sampling frame or from external data sources (Groves, 2006; Schouten et al., 2009; Montaquila and Brick, 2009; Wagner, 2010; Andridge and Little, 2011; 2020; Särndal and Lundquist, 2014; Brick and Tourangeau, 2017; Hedlin, 2020). Results may vary depending on the choice of adjustment approaches for nonresponse.

Responding to a solicitation from the U.S. National Center for Education Statistics, we develop exemplars to help guide practices of the nonresponse bias analysis (NRBA) for educational assessment surveys. This article describes an exemplar developed on the Early Childhood Longitudinal Study, Kindergarten Class of 2010-11 (ECLS-K:2011) (NCES, 2013). The ECLS-K:2011 study collects national data on children as they progress from kindergarten through the 2015–16 school year, when most of them will be in fifth grade. The ECLS-K:2011 program is unprecedented in its scope and coverage of child development, early learning, and school progress, drawing together information from multiple sources, including school administrators, parents, teachers, early care and education providers, and children.

We focus here on a cross-sectional NRBA based on the first wave, 2010 fall data collection. Nonresponse in the ECLS-K occurs when schools in the sample are missing due to lack of cooperation and when children, parents, and teachers are nonrespondents within schools. We summarize the current NRBA implementation in the ECLS-K study, which has the limitation of the strong reliance on the missing at random (MAR) assumption (Rubin, 1976) and is described formally in Section 5. A major objective of our exemplar is the systematic formulation of NRBA steps to guide practice. These steps include a sensitivity analysis based on proxy pattern-mixture models (Andridge and Little, 2011; 2020), which allows for missing not at random (MNAR) missingness mechanisms. We present the NRBA measures and evaluate the quality of such measures based on the predictive performances of auxiliary variables in multivariate models.

The paper is organized as below. Section 2 discusses the background of missing data and NRBA. Section 3 describes the sensitivity analysis framework. The systematic NRBA with ten detailed steps is summarized in Section 4. We demonstrate the NRBA steps with the ECLS-K:2011 study in Section 5. Section 6 concludes with challenges and future extensions.

## 2. Background

The *pattern* and *mechanism* of missing data play an important role informing the potential bias from unit or item nonresponse. The pattern refers to which values in the data set are observed and which are missing. Specifically, let $Y = (y_{ij})$ denote an $(n \times p)$ rectangular dataset without missing values, with $i$th row $y_i = (y_{i1}, \ldots, y_{ip})$ where $y_{ij}$ is the value of $j$th variable $Y_j$ for subject $i$, where $i = 1, \ldots, n$ total number of subjects and $j = 1, \ldots, p$ total number of variables. With missing values, the pattern of missing data is defined by the *response indicator matrix* $R = (r_{ij})$, such that $r_{ij} = 1$ if $y_{ij}$ is observed and $r_{ij} = 0$ if $y_{ij}$ is missing; equivalently, $1 - r_{ij}$ is the *missing-data indicator* for $y_{ij}$.

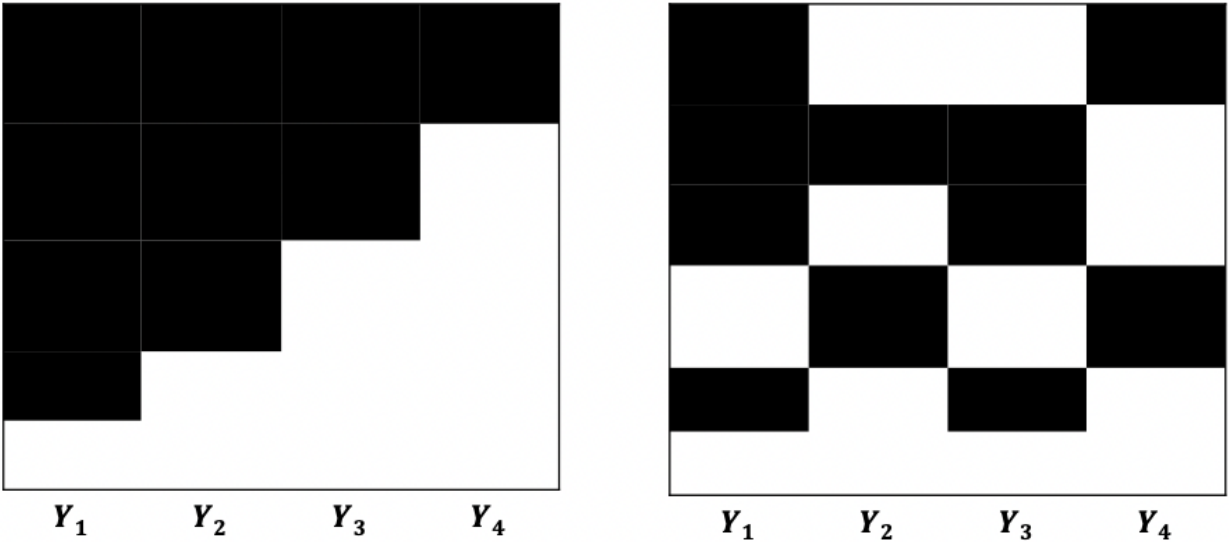

**Figure 1. Illustration of missing patterns and the implied response indicator matrices with four variables. (black areas indicate response with $R = 1$, and white areas indicate nonresponse with $R = 0$). The left plot is a monotone "staircase" pattern, and the right shows a general "swiss-cheese" pattern of missingness.**

Unit nonresponse occurs when the survey variables $Y$ are missing for units subject to nonresponse and leads to a special case of the *monotone* missing data, where the variables can be ordered so that $Y_{j+1}, \ldots, Y_p$ are missing for all subjects where $Y_j$ is missing, for all $j = 1, \ldots, p - 1$, illustrated in the left plot of Figure 1. The monotone pattern often arises in longitudinal data subject to attrition, where once an individual drops out, no more data are observed for that person. In the cross-sectional ECLS-K data, the individual units of children will be missing if their school is missing, and a monotonic pattern arises for student-level and school-level data. Because student-level data are not observed for schools that do not respond to the survey, as an illustration, we can use $Y_1$ to denote the collected school characteristics and $Y_2$ to denote the children assessment variables. Item nonresponse occurs when the unit only responds to partial survey measures, and the right plot of Figure 1 shows that item nonresponse leads to a *general* "swiss-cheese" pattern of missingness.

The missingness mechanism addresses why values are missing and whether these reasons relate to values in the data set. For example, schools or pupils with schools that refuse to participate in the ECLS-K may differ in academic performances from schools or pupils that participate. Rubin (1976) treats $R$ as a random matrix and characterizes the missingness mechanism by the conditional distribution of $R$ given $Y$, say $f(R|Y,\psi)$, where $\psi$ denotes unknown parameters. Let $Y_{(1)}$ denote the observed components of $Y$, $Y_{(0)}$ denote the missing components of $Y$, and $X$ be the set of variables observed for respondents and nonrespondents. When missingness does not depend on the values of the data $X$ or $Y$, missing or observed, that is,

$$f(R|X,Y,\psi) = f(R|\psi) \text{ for all } Y, \psi,$$

the missingness is called missing completely at random (MCAR), and MCAR missing data lead to an increase in the variance of estimates but do not lead to bias. However, MCAR is a strong and unrealistic assumption in most survey settings, including the ECLS-K.

A less restrictive assumption is that missingness depends only on $X$ and the values $Y_{(1)}$ that are observed, and not on values $Y_{(0)}$ that are missing. That is, if $Y_{(0)}, Y_{(0)}^*$ are any two sets of values of the missing data, then:

$$f(R|X, Y_{(1)}, Y_{(0)}, \psi) = f(R|X, Y_{(1)}, Y_{(0)}^*, \psi) \text{ for all } X, Y_{(0)}, Y_{(0)}^*, \psi. \tag{1}$$

The missingness is called MAR at the observed values of $R$ and $Y_{(1)}$. If (1) does not hold, the data are MNAR.

Most existing analyses, either by nonresponse weighting or imputation, make the MAR assumption, in part because MNAR analyses are often strongly reliant on untestable assumptions. The inclusion of variables predictive of survey variables strengthens the NRBA, providing more confidence that the MAR assumption is justified. In contrast, an NRBA based on variables that are weakly related to key survey variables provides weak evidence pro or con nonresponse bias; in other words, lack of evidence of bias from such an analysis does not imply lack of bias.

A simple expression of nonresponse bias for the mean of respondents $\bar{y}_R$ is

$$Bias(\bar{y}_R) = \frac{N - N_R}{N} (\bar{Y}_R - \bar{Y}_{NR}), \tag{2}$$

where $N$ is the population size, $N_R$ is the number of respondents, $\bar{Y}_R$ and $\bar{Y}_{NR}$ are the respondent and nonrespondent means in the population, respectively. The bias is zero if missingness is MCAR, but that is generally a strong and untenable assumption.

If variables are measured for respondents and nonrespondents, either survey variables measured on other levels in the sample (e.g., collected school characteristics for nonresponding students) or auxiliary variables from the sampling frame and a census or large survey of the population, they can be used to attempt to reduce bias. The main approaches to bias adjustment are *nonresponse weighting*, where responding units are weighted by the inverse of an estimate of the probability of response, and *imputation*, where missing values are imputed by predictions based on observed variables. Weighting is commonly used to adjust for unit nonresponse, and imputation is usually applied to handle item nonresponse because it is more effective than weighting for handling general patterns of missing data (Little and Rubin, 2019).

To construct response propensity weights for unit nonresponse, let $R_j$ be the indicator for response to $Y_j$, for $j = 1, 2$. With a monotone pattern of data with $Y_2$ less observed than $Y_1$, the MAR condition for missingness of $Y_2$ can be weakened to:

$$Pr(R_2 = 1|X, Y_1, Y_2) = Pr(R_2 = 1|X, Y_1). \qquad (3)$$

That is, the propensity to respond to the survey variable $Y_2$ can depend on the values of survey variables $Y_1$. Assuming MAR with the monotone pattern, $Pr(R_1 = 1|R_2 = 1) = 1$, the probability of response to $Y_2$ can be factored as

$$Pr(R_2 = 1|X, Y_1) = Pr(R_1 = R_2 = 1|X, Y_1) = Pr(R_1 = 1|X) \times Pr(R_2 = 1|R_1 = 1, X, Y_1), \qquad (4)$$

and the conditional probability of $R_2$ given $R_1$ can depend on values of $Y_1$ as well as $X$. The response weight for variable $Y_j$ is then the product up to variable $Y_j$ of the inverse of these estimated conditional propensities (Little and David, 1983). In the ECLS-K study, the unit refers to a student. When the school refuses to participate in the study, all students in that school will be missing; And in participating schools only a subset of students respond. Therefore, the school-level and child-level missing-data patterns are monotonic, where $R_1$ denotes the school-level response indicator and $R_2$ denotes the child-level response indicator. Applying this factorization in (4) to the ECLS-K study, we model the conditional response propensity of children given the observed variables $X$ and school characteristics $Y_1$: $Pr(R_2 = 1|R_1 = 1, X, Y_1)$.

**Table 1. Bias and Variance of Multiple Imputation (MI) and Inverse Propensity Weighting (IPW) Relative to Complete Case (CC) analysis for Estimating a Mean, by Strength of Association of the Auxiliary Variables with Response (R) and Outcome (Y) (Little, Carpenter and Lee, 2022).**

| Association of X with Response R | Association of X with Outcome Y | | | |
|---|---|---|---|---|
| | Propensity: Low<br>Other Xs: Low | Propensity: Low<br>Other Xs: High | Propensity: High<br>Other Xs: Low | Propensity: High<br>Other Xs: High |
| Low | Cell LLL<br>　　IPW　MI<br>Bias:　---　---<br>Var:　---　--- | Cell LLH<br>　　IPW　MI<br>Bias　---　---<br>Var　---　↓ | Cell LHL<br>　　IPW　MI<br>Bias:　---　---<br>Var:　↓　↓ | Cell LHH<br>　　IPW　MI<br>Bias:　---　---<br>Var:　↓　↓↓ |
| High | Cell HLL<br>　　IPW　MI<br>Bias:　---　---<br>Var:　↑　--- | Cell HLH<br>　　IPW　MI<br>Bias:　---　---<br>Var:　↑　↓ | Cell HHL<br>　　IPW　MI<br>Bias:　↓　↓<br>Var:　↓　↓ | Cell HHH<br>　　IPW　MI<br>Bias:　↓　↓<br>Var:　↓　↓↓ |

Notes:

For characterizing the association between X and Y, the X is split into the propensity, that is the best predictor of R in the regression of R on X, i.e., Propensity, and components of X orthogonal to the propensity, i.e., Other Xs. The two columns represent two types of association between X and Y are distinguished, the strength of association between the propensity to respond and Y, and the strength of association between other Xs and Y, respectively. With a single X, the propensity is a function of X and other X is a null set.

"CC" for complete case analysis, "IPW" for inverse propensity weighting, and "MI" for multiple imputation; "---" for Bias (or Var) within a cell indicates that the estimate for the method has similar bias (or variance) to the estimate for CC; "↓" for Bias (or Var) within a cell indicates that the estimate for the method has less absolute bias (or variance) than the estimate for CC; "↓↓" for Bias (or Var) within a cell indicates that the estimate for the method has much less absolute bias (or variance) than the estimate for CC; "↑" for Bias (or Var) within a cell indicates that the estimate for the method has greater absolute bias (or variance) than the estimate for CC. In summary, "↓" indicates that a method is better than CC, "↑" indicates that a method is worse than CC, and "---" indicates that a method is similar to CC.

To handle item nonresponse, a drawback of single imputation is that it overestimates precision of survey estimates. A recommended solution to this problem is *multiple imputation* (MI), where missing values are drawn from their predictive distributions, and multiple data sets are created with different draws of the missing data imputed. Although the theories are rooted in Bayesian statistics, MI can be applied with replication sampling methods, such as bootstrap and jackknife algorithms, to take into account imputation uncertainty. Estimates of the resulting data sets are then combined using Rubin's MI combining rules. See, for example, Rubin (1987) or Little and Rubin (2019). A useful feature of MI for practitioners is that a wide variety of software for MI is now available, as summarized by Yucel (2011) and Si et al. (2021).

To reduce bias, auxiliary variables in the nonresponse adjustment must be predictive of both the survey variable of interest and nonresponse indicator (Little and Vartivarian, 2005). Table 1 presents the bias and variance of MI and inverse propensity weighting for estimates of means, compared to unadjusted analyses based on the complete cases. The variances are calculated based on large-sample approximations. Taken from Little et al. (2022), this table is a refinement of the simpler table in Little and Vartivarian (2005). Weighting is only effective when the auxiliary variables are related to the survey variables; otherwise, it increases the variance with no reduction in bias (Little and Vartivarian, 2005; Little et al., 2022).

3. **Methods for nonresponse bias analysis and sensitivity analysis**

A substantial difference between unadjusted estimates and estimates adjusted by weighting or imputation suggests that nonresponse adjustment is important, and hence is often a component of NRBA. The key to a useful NRBA is to identify a rich set of auxiliary variables $X$ that are highly predictive for the survey variables $Y$. These might include variables in the sampling frame for bias adjustments, from external data sources that are not included in the analysis of the data, and also available via data linkage.

We fit multivariate models for the survey variable $Y$ and response propensity $\Pr(R = 1)$ and obtain the predicted values for both respondents and nonrespondents. We fit a multivariate regression model with all the auxiliary variables because the variables are often adjusted simultaneously, and checking marginal relationships cannot account for the correlation among auxiliary variables. Variable selection procedures, such as the stepwise forward selection and LASSO (Tibshirani, 1996), can be implemented to select predictive variables and handle multicollinearity among a large number of auxiliary variables. To assess the correlation between the survey variable and response propensity, we group respondents into strata based on the quintiles of the predicted response propensities, $\hat{P}_r(R_i = 1)$, and compare the distributions of survey variables across subgroups. To compare the mean differences of the survey variable between respondents and nonrespondents, we conduct sensitivity analyses under different missing data mechanisms. Conditioning on auxiliary variables and observed survey variables, we denote the predictions of Y as a proxy variable X, where $X$ is available for both respondents and nonrespondents. Here the proxy variable $X$ is a summary of available information denoted by $(X, Y_1)$ to predict $Y_2$ in Equations (3) and (4). We fit proxy pattern-mixture models for the distribution of $(X, Y, R)$ in the population (Little 1994; Andridge and Little 2011; Little et al., 2020): $f(X, Y, R) = f(X, Y|R)f(R)$, where the joint distribution of $(X, Y)$ varies between the respondents $(R = 1)$ and nonrespondents $(R = 0)$, specified as below.

$$f(X,Y|R = r) = \text{Bivariate} - \text{Normal}\left[\begin{pmatrix}\mu_x^{(r)}\\\mu_y^{(r)}\end{pmatrix}, \begin{pmatrix}\sigma_{xx}^{(r)} & \sigma_{xy}^{(r)}\\\sigma_{xy}^{(r)} & \sigma_{yy}^{(r)}\end{pmatrix}\right], \quad r = 0 \text{ or } 1,$$

(5)

$$Pr(R = 1|X,Y) = g(V), \text{ where } V = (1 - \phi)\sqrt{\frac{s_{yy}}{s_{xx}}}X + \phi Y,$$

which is a bivariate-normal distribution with mean $(\mu_x^{(r)}, \mu_y^{(r)})$ and variance-covariance parameters $(\sigma_{xx}^{(r)}, \sigma_{xy}^{(r)}, \sigma_{yy}^{(r)})$, and the constraint $g(V)$ is a link function (e.g., logit or probit) of $(X,Y)$, the pre-specified constant $\phi$, and the estimated sample variance for respondents based on observed data, $s_{yy} = \sigma_{yy}^{(1)}, s_{xx} = \sigma_{xx}^{(1)}$. This additive assumption of $(X,Y)$ in $g(V)$ requires that the effect of X on the missingness is not moderated by Y. Since X is the best prediction of Y given the observed variables, we assume that the proxy variable X and the survey variable Y are positively correlated.

We can estimate the mean $\overline{Y}$ as $\hat{\mu}_y$ based on the proxy pattern mixture model. The NRBA index is the difference between $\hat{\mu}_y$ and the respondent mean $\overline{y}_R$ of $\overline{Y}$,

$$NRBA(\phi) = \hat{\mu}_y - \overline{y}_R = g(\hat{\rho}, \phi)\sqrt{\frac{s_{yy}}{s_{xx}}}(\overline{x} - \overline{x}_R)$$

(6)

$$= \frac{\phi + (1 - \phi)\hat{\rho}}{\phi\hat{\rho} + (1 - \phi)}\sqrt{\frac{s_{yy}}{s_{xx}}}(\overline{x} - \overline{x}_R),$$

where $\overline{x}_R$ is the respondent mean of X, and $\overline{x}$ is the sample mean of X. The function $g(\hat{\rho}, \phi) = \frac{\phi+(1-\phi)\hat{\rho}}{\phi\hat{\rho}+(1-\phi)}$ is a function of the respondent sample correlation $\hat{\rho}$ of X and Y and a sensitivity parameter $\phi$, $0 \leq \phi \leq 1$. Here $g(\hat{\rho}, \phi)$ increases with the strength of the proxy, i.e., $g(\hat{\rho}, \phi) \to 1$ as $\hat{\rho} \to 1$. With $\phi = 0$, $g(\hat{\rho}, 0) = \hat{\rho}$, the nonresponse is MAR. We will try different values of $\phi$ and compare the effects on the mean estimates. The NRBA index requires the calculation of: 1) $\hat{\rho}$, the correlation between the proxy and survey variables, and 2) $d = \bar{x} - \bar{x}_R$, the mean differences of the proxy variable between the population and the respondents. Andridge and Little (2011) have discussed the effect of different correlation $\hat{\rho}$ values on the NRBA. Moderate values of correlation $\hat{\rho}$, such as 0.5, and even low correlation can provide useful evidence. The choices of the threshold values for both the correlation $\rho$ and the difference $d$ have to depend on the substantive application and whether they alter the key findings. As a subjective recommendation, we would suggest that a correlation of less than 0.4 is weak, a correlation from 0.4 to 0.7 is moderate, and a correlation of more than 0.7 is strong. We consider a difference of less than $0.1 * s_{xx}$ as small, between $0.1 * s_{xx}$ and $0.3 * s_{xx}$ as medium-range, and larger than $0.3 * s_{xx}$ as large. In a survey with multiple outcome measures of interest, we can use the ranking based on the list of $\hat{\rho}$ and $d$ values in the NRBA.

When the inferential interest is subgroup analysis, for example, educational assessments across different race/ethnicity groups, we modify the expression (6) for each subgroup $k$, for $k = 1, \ldots K$, and obtain the subgroup mean estimates,

$$NRBA_k(\phi) = \hat{\mu}_{yk} - \bar{y}_{Rk} = \frac{\phi + (1-\phi)\hat{\rho}_k}{\phi\hat{\rho}_k + (1-\phi)} \sqrt{\frac{s_{yy.k}}{s_{xx.k}}} (\bar{x}_k - \bar{x}_{Rk}). \tag{7}$$

Here $\hat{\rho}_k$ is the correlation between $(X, Y)$, $s_{xx.k}$ and $s_{yy.k}$ are the respondent sample variances of $X$ and $Y$, in subgroup $k$. West et al. (2021) develop regression coefficient estimates with the proxy pattern-mixture models and generate implicit subgroup mean estimates. We extend their results and consider group-specific correlation and variance values. The underlying model we consider adds interactions between subgroup indicators and auxiliary variables in the model for $Y$, while the model in West et al. (2021) only includes main effects and results in the same values of partial correlation and variances across subgroups.

## 4. Main steps of the nonresponse bias analysis

Figure 2 summarizes the steps of our proposed systematic NRBA.

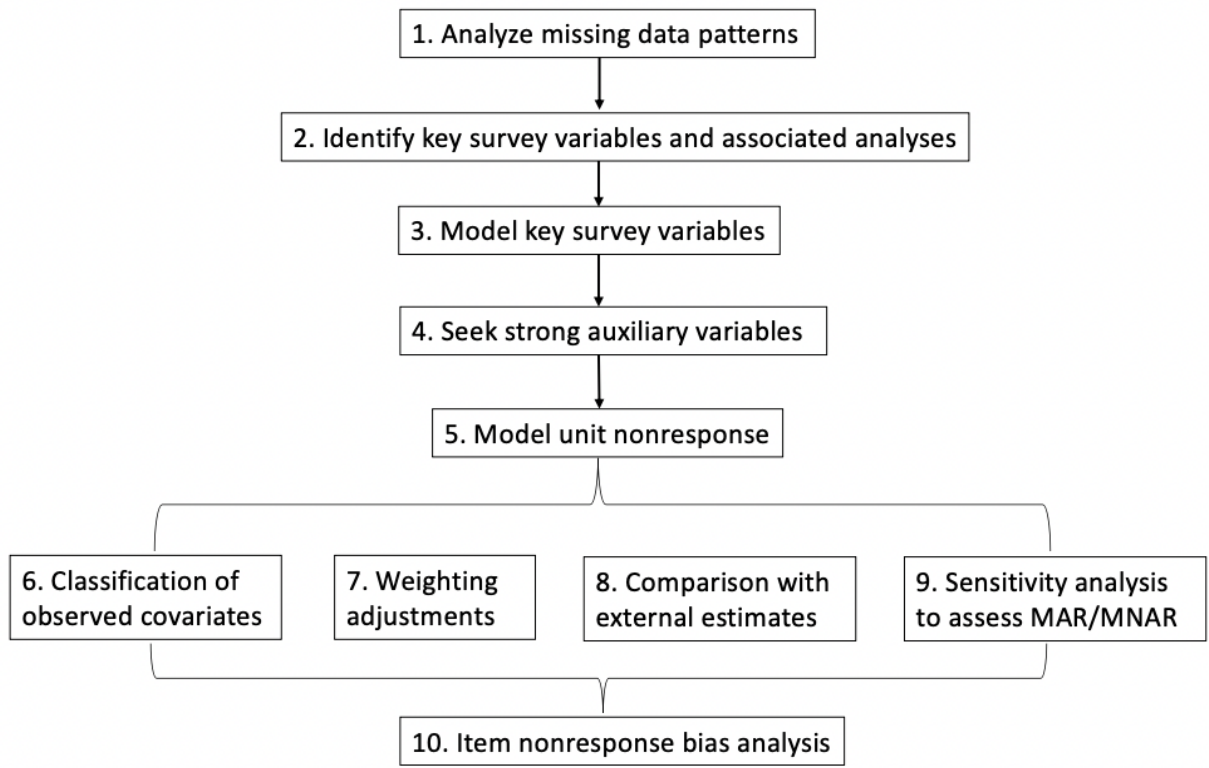

**Figure 2. Nonresponse bias analysis process.**

1. **Analyze missing-data patterns.** Describe the missingness proportions of individual variables and the missingness pattern across variables. The size of bias is likely to be related to the extent of missing data. More generally, missingness patterns can indicate variables or sets of variables with high levels of nonresponse, which are likely to be vulnerable to nonresponse bias.

2. **Identify key survey variables and associated analyses.** Generally, surveys measure a large number of variables, and it is not feasible to include them all in an NRBA. Thus, identify a small set of "key" survey variables that represent the main subject-matter content of the survey. Also, identify several analyses of interest involving these variables. These could be descriptive or analytic in nature. For survey analysts, such as educational researchers, the NRBA will focus on the specific research questions of interest.

3. **Model key survey variables as a function of fully observed predictors.** The key to a successful NRBA is to find and include "strong" variables that are observed for respondents and nonrespondents and are predictive of key survey variables. Thus, key survey variables should be regressed on fully observed variables to identify predictors of one or more of the key survey variables.

4. **Seek strong observed predictors in auxiliary data.** Many existing NRBAs are limited by the absence of such variables in the data set (Kreuter et al. 2010). Particularly if the analysis in Step 3 indicates that strong variables are absent in the dataset, attempt to link the survey data to external information containing auxiliary variables observed for both respondents and nonrespondents and predictive of survey variables. Such variables are useful for the NRBA, whether measured on individuals in the survey or in aggregate forms, such as marginal proportions or means.

5. **Model unit nonresponse as a function of observed predictors.** Variables that are strongly related to nonresponse are important for nonresponse bias adjustment to the extent that they are also predictive of survey variables. Predictors that are weakly related to nonresponse but strongly related to survey variables do not lead to bias adjustment but can improve the precision of survey estimates (e.g., Little and Vartivarian, 2005). If external variables are identified in Step 4 and measured at the unit level, separate models should be developed for (a) variables restricted to those included in the database and (b) variables including external variables identified in Step 4.

6. **Assess observed predictors for the potential for bias adjustment.** Results from Steps 2-5 provide the basis for the classification of observed predictors according to the eight cells of Table 1. Key variables for nonresponse bias adjustment are predictive of both nonresponse and one or more key survey variables (Little and Vartivarian, 2005).

7. **Assess the effects of nonresponse weighting adjustments on key survey estimates.** Based on both statistical inferences and substantial findings, small changes between unweighted and weighted estimates suggest that bias may be small, particularly if the adjustment is based on variables that are strongly related to survey variables of interest, as identified in Step 4.

8. **Compare the survey with external data using summary estimates of key survey variables**. Ideally, the external data should be of high quality and are close to serving as the proxy of true values. The comparisons inform potential nonresponse bias, but differences in the estimates could be due to other sources of heterogeneity.

9. **Perform a sensitivity analysis to assess the impacts of deviations from MAR.** Based on the NRBA measure in Equation (6) or (7), we recommend trying different values of the sensitivity parameter and comparing the estimates under different missingness mechanisms. If the resulting confidence intervals largely overlap, the estimates are not sensitive about MAR assumptions. The correlation $\hat{\rho}$ between the proxy and survey variables indicates the quality of the NRBA measure, where larger $\hat{\rho}$ means stronger evidence.

10. **Conduct item nonresponse bias analyses for all analyzed variables.** As with unit nonresponse, the size of bias is related to the amount of missing information. Item nonresponse often results in general missing-data patterns, and an assessment of the degree of item nonresponse can be obtained from Step 1. Item nonresponse bias analysis using the above steps may be indicated for key survey items with high levels of item nonresponse. The values of items that are fully or close to fully observed can be included as additional predictors in item nonresponse models, and MI software for general patterns of missingness allows for fully exploiting information on the observed items. To yield valid inferences of the quantities of interest, imputation approaches that take into account the correlation structure with available data and propagate uncertainty due to missing data improve bias reduction and estimation precision (Si et al., 2021).

5. ECLS-K analysis: background and application

We demonstrate the ten steps of our proposed NRBA in the ECLS-K:2011 study with 15,830 responding students out of 18,170 total eligible units. An NRBA needs to account for the design of the survey. We briefly introduce the ECLS-K:2011 sampling design, weighting adjustment, and current NRBA procedures.

The ECLS-K:2011 adopts a three-stage sample design, with geographic areas as primary sampling units (PSUs), schools sampled within PSUs, and children sampled within schools. A stratified sample of PSUs is selected with probability proportional to size (PPS), where the measure of size is the estimated number of 5-year-old children in the PSU, with oversampling of Asians, Native Hawaiians, and other Pacific Islanders (APIs). All PSUs are grouped into 40 strata defined by metropolitan statistical area status, census geographic region, size class (defined using the measure of size), per capita income, and the race/ethnicity of 5-year-old children residing in the PSU. The sources for the school frames are the most recent Common Core of Data (2006–07 CCD) and the Private School Survey (2007–08 PSS). Schools are selected with PPS. The measure of size for schools is kindergarten enrollment adjusted to account for the desired oversampling of APIs. Schools are also sampled from the supplemental frame of newly opened schools and added kindergarten programs that are not in the original frames, and the selection probability for a new school in an existing PSU is conditional on the within-stratum probability of selecting that PSU. Public school substitution is conducted in nonparticipating districts assigned with the base weight of the original school, adjusted for school size differences. In the third stage of sampling, children enrolled in kindergarten of graded schools and 5-year–old children in ungraded schools are selected within each sampled school. Two independent sampling strata are formed within each school, one containing API children with a higher sampling rate than the second containing all other children. Within each stratum, children are selected using equal probability systematic sampling and the target number of 23 at any one school. Once the children are sampled from the school lists of enrolled kindergartners, parent contact information for each child is obtained from the school. The information is used to locate a parent or guardian, conduct the parent interview, and gain parental consent for the child to be assessed. Teachers who teach the sampled children and before- and after-care providers are also included in the study and asked to complete questionnaires.

For the base year of ECLS-K, weights are provided at the child and school levels as the inverse of the probability of the multi-stage selection. The ECLS-K applies raking to external control margins (Deming and Stephan, 1940). The base-year coverage-adjusted child base weight is raked to external control totals from the number of kindergartners enrolled in public schools in the 2009-10 CCD and in private schools in the 2009-10 PSS, the two most up-to-date school frames available at the time of weight computations that are also the closest to the time frame of the kindergarten year of the ECLS-K:2011. Raking cells are created using census region (Northeast, Midwest, South, West), locale (city, suburb, town, rural), school type (public, Catholic, non-Catholic private, non-religious private), and kindergarten size (fewer than 85 and 85 or more). After raking, the extremely large weights are trimmed.

The response status is used to adjust the base weight for nonresponse to arrive at the final full sample weight. Nonresponse classes are formed separately for each school type (public/Catholic/non-Catholic private). Within school type, analysis of child response propensity is conducted using child characteristics such as date of birth and race/ethnicity to form nonresponse classes. The child-level nonresponse adjustment factor is computed as the sum of the weights for all the eligible (responding and nonresponding) children in a nonresponse class divided by the sum of the weights of the eligible responding children in that nonresponse class.

An NRBA of ECLS-K:2011 by Westat (Tourangeau et al., 2013) examines unit nonresponse with four approaches. The first approach reports school-level and student-level response rates for subgroups--an analysis related to Steps 1 and 5 above. The response rates show variation across school types, census regions, locale, kindergarten enrollment, percent minority, race/ethnicity, and years of birth, and large variation increases the potential for nonresponse bias. With a similar role, the R indicator (Schouten et al., 2009) measures the variability in the probability of responding to a survey as a function of auxiliary variables. Response rates and R indicators are agnostic with regard to specific survey variables of interest, failing to reflect the fact that selection bias depends on the strength of the relationship of selection with the survey variable. We recommend in Step 2 to select a few key variables of interest, which are child assessment outcomes in the ECLS-K:2011 study. The second and third approaches compared sample estimates to estimates computed from the sampling frame, the Census data, and other sources, similar to our recommendation in Step 8. The fourth approach compares ECLS-K:2011 estimates weighted with and without nonresponse adjustments, as recommended in Step 7. Larger differences could be indicative of substantial nonresponse bias; however, the strength of this evidence depends on whether the characteristics used in the nonresponse adjustment are strongly related to survey variables of interest.

Our proposed NRBA process distinguishes from the current practice (Tourangeau et al., 2013) with three aspects: 1) We explicitly conduct an outcome-specific NRBA and examine multiple key survey variables; 2) We calculate the NRBA measures and evaluate the quality of such measures based on the predictive performances of auxiliary variables in multivariate models; and 3) We conduct sensitivity analyses to assess the impact of deviations from MAR. The multi-stage sampling of PSUs, schools, and children results in nonresponse for schools and children, and there are school substitutes to replace the nonresponding schools. As an illustration, we focus on the child-level NRBA with interest in estimating the mean values of child assessment outcomes overall and across subgroups of interest.

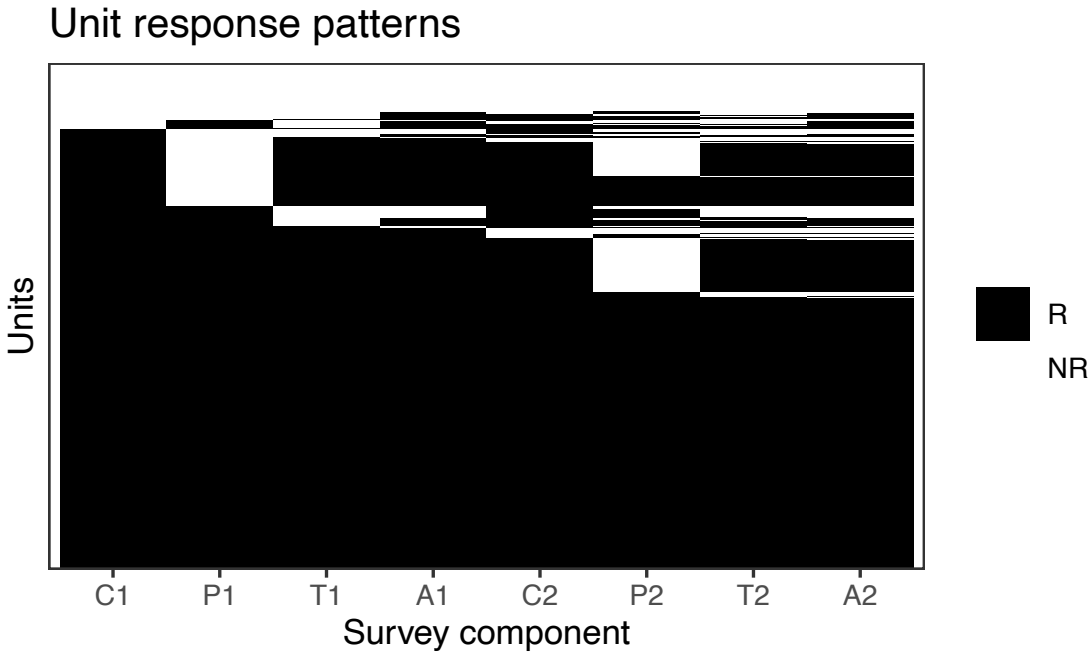

**Figure 3. Unit response (R marked by black areas) and nonresponse (NR marked by white areas) patterns of child (C), parent (P), teacher (T) and teacher student-level assessment (A) survey instruments for fall 2010 (1) and spring 2011 (2) data collection. C1: child survey in fall 2010; P1: parent survey in fall 2010; T1: teacher survey in fall 2010; A1: teacher-student assessment survey in fall 2010; C2: child survey in spring 2011; P2: parent survey in spring 2011; T2: teacher survey in spring 2011; and A2: teacher-student assessment survey in spring 2011.**

SOURCE: U.S. Department of Education, National Center for Education Statistics, Early Childhood Longitudinal Study, Kindergarten Class of 2010-11, 2010 Fall and 2011 Spring.

### 5.1. Step 1: Analyze missing-data patterns.

The school-level and child-level missing-data patterns are monotonic, shown in the left plot of Figure 1, so school characteristics can be used in the child-level NRBA. In Figure 3, we present the unit nonresponse patterns of the child, parent, teacher, and teacher-student assessment surveys both for the fall and spring collection. The goal is to check response rates and whether nonrespondents in one variable could have information available from other variables that can be used in the NRBA. Conditional response propensities can be computed based on the factorization given in Equation (4).

Detailed response rates by school/child characteristics are reported in the User Manual of the ECLS-K:2011 Kindergarten Study (Tourangeau et al., 2013) and generally high during the base year, with an overall rate of 69% for schools, 87% for children, 74% for parents, and 82% for teachers. Looking into the missing data patterns of Figure 3, we do not have information on parents or teachers for most of the nonresponding children and for a small proportion of responding children. Characteristics of child respondents could be useful to inform the NRBA of parent and teacher interviews.

## 5.2. Step 2: Identify key survey variables and associated analyses.

Since the ECLS-K study focuses on child development, we have identified a few key survey variables on child assessment outcomes: reading scores estimated by the item response theory (IRT) (Hambleton et al., 1991), mathematics IRT scores, child body mass index (BMI), scores on being impulsive/overactive and self-control based on parent interviews, and scores on externalizing and internalizing problems based on teacher interviews.

We are interested in estimates of means in the population and in mutually exclusive subgroups defined by race/ethnicity (Non-Hispanic White, Non-Hispanic Black, Hispanic, API, and Other) and school type (public, private).

## 5.3. Step 3: Model key survey variables as a function of fully observed predictors.

For the cross-sectional NRBA at the baseline, we have frame variables from the 2006-07 CCD for public schools and the 2007-08 PSS for private schools. We include the public and private schools and exclude schools selected from the supplemental frames. Because the sample size of private schools is smaller than that of public schools, we conduct the NRBA by combining the two sampling frames. That is, we identify overlapping variables between the CCD and PSS data and select one set of frame variables that are available for both respondents and nonrespondents in the sample.

The auxiliary variables include sex, year of birth, race/ethnicity, school type, census region, locale, the number of students, number of full-time-equivalent teachers, student to teacher ratio, lowest and highest grades offered, percentages of kindergarteners, American Indians, Asians, Hispanics, and Black in schools.

Given the auxiliary variables available for both respondents and nonrespondents, the survey variable is conditionally independent of the response indicator. Because the survey variables are continuous, we fit a linear regression model for $Y$ given the auxiliary variables. As alternatives to regression models, tree-based models, random forests, and gradient boosting algorithms can be used to model the survey variable, as well as the response propensity. Machine learning algorithms automatically detect interactions and nonlinear relationships and could yield good predictive performances. The nodes determined by a tree structure can be used as weighting classes for nonresponse adjustments. As an illustration, we use tree-based models for variable selection and regression models for prediction. We fit a tree model to select high-order interaction terms. Then we include all main effects and the identified interactions into the linear regression for $Y$ and perform a stepwise variable selection based on the Akaike information criterion to determine the final models.

Using the reading IRT score as an example, the selected predictors in the final outcome model include race/ethnicity, year of birth, sex, school locale, region, lowest and highest grades offered, school type, the number of enrolled students, the number of full-time-equivalent teachers, percentages of Hispanics, Asian, and Black in school, the two-way interactions between locale and school type, between locale and the percentage of Asian, between race/ethnicity and region, between race/ethnicity and the percentage of Hispanics, between race/ethnicity and the percentage of Black, between race/ethnicity and the lowest grades offered, and between locale and the number of full-time-equivalent teachers.

We predict the outcome values for both respondents and nonrespondents and obtain the proxy variable $X$. The correlation between the outcome $Y$ and the proxy variable $X$ for the respondents is $\hat{\rho} = 0.36$. Hence, $X$ is a moderately weak proxy with small $\hat{\rho}$.

**5.4. Step 4: Seek strong observed predictors in auxiliary data.**

The analysis in Step 3 indicates that strong variables are absent in the dataset, and efforts should be made to add more predictors in the model for $Y$ and improve the prediction.

For the child assessment outcomes, the highly predictive variables include poverty level, socioeconomic status, the type of language use at home, parental education, parental marital status at the time of birth, and nonparental care arrangements during the year prior to kindergarten. The correlation $\hat{\rho}$ between the survey variable $Y$ and the proxy variable $X$ increases to 0.48 after adding them to the outcome model. However, they are only available for a small proportion of nonrespondents, 640 out of 2320 nonrespondents having such information.

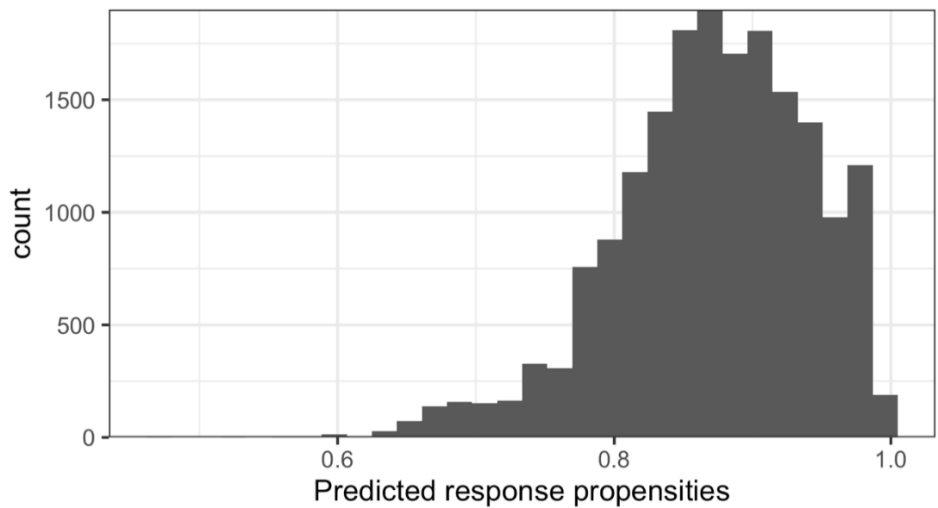

**Figure 4. Frequency distribution of the predicted response propensities by the logistic regression.**

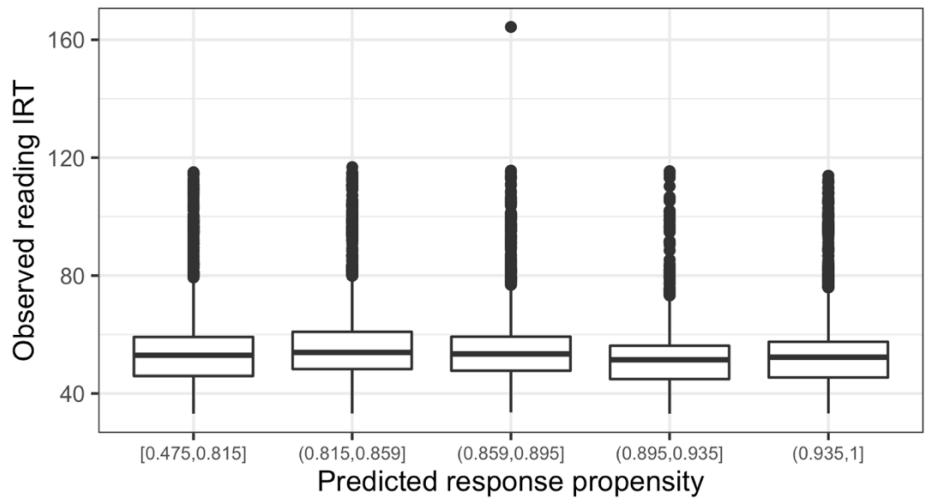

**Figure 5. Reading score distributions of respondents stratified by the quintiles of predicted response propensities.**

SOURCE: U.S. Department of Education, National Center for Education Statistics, Early Childhood Longitudinal Study, Kindergarten Class of 2010-11. 2010 Fall.

**5.5. Step 5: Model unit nonresponse as a function of observed predictors.**

First, we model the school-level response propensities with logistic regression and find that the predictive variables include school type and percentages of kindergarteners and Asians in schools. The model for the school-level response propensities has a value of 0.60 for the Area Under the receiver-operating characteristic Curve (AUC), an assessment of discriminatory ability. The small AUC value shows that the auxiliary variables are weakly predictive of the school-level response.

Next, we use a response indicator with $R_i = 1$ if the child $i$ responds to the study; otherwise $R_i = 0$. Including the school-level characteristics and auxiliary variables that are available for both responding and nonresponding children as predictors, we fit a logistic regression with the children response indicator $R_i$ as the outcome to estimate the conditional child-level response propensity. The final model has an AUC value of 0.67. The selected predictors include race/ethnicity, region, locale, the number of enrolled students, number of full-time-equivalent teachers, student to teacher ratio, highest grades offered, percentages of Hispanics and Black in schools, the two-way interaction between race/ethnicity and region, and the two-way interaction between race/ethnicity and the percentage of Black students in the school.

Figure 4 depicts the frequency distribution of the predicted child-level response propensities $\widehat{Pr}(R_i = 1)$ by the logistic regression. The predicted values are generally large, as the overall response rate is high (87%) and presents a modestly small amount of variation. We examine the relationships between predicted response propensities and key survey outcomes based on the respondents. Figure 5 presents the reading score distributions of respondents stratified by the quintiles of predicted response propensities. The boxplots show that the outcome distributions do not change across response propensity strata. That is, the outcome is weakly correlated with the response propensity, and the estimated correlation is -0.05.

**5.6. Step 6: Assess observed predictors for the potential for bias adjustment.**

**Table 2. Comparison of unweighted, nonresponse-unadjusted weighted estimates and weighted estimates (SE: standard errors; Deff: design effects).**

|  | Unweighted | | Nonresponse-unadjusted, base weighted | | Nonresponse-adjusted, weighted | | |
|---|---|---|---|---|---|---|---|
|  | mean | SE | mean | SE | mean | SE | Deff |
| Overall | 54.07 | 0.30 | 53.89 | 0.30 | 53.85 | 0.31 | 11.68 |
| Race/ethnicity | | | | | | | |
| White (not Hisp) | 55.45 | 0.34 | 55.54 | 0.36 | 55.54 | 0.35 | 7.62 |
| Black (not Hisp) | 52.25 | 0.58 | 52.26 | 0.48 | 52.17 | 0.48 | 5.13 |
| Hispanic | 50.32 | 0.49 | 50.19 | 0.49 | 50.13 | 0.49 | 10.05 |
| API (not Hisp) | 58.17 | 0.88 | 57.20 | 1.00 | 57.37 | 1.04 | 6.74 |
| Other | 56.25 | 0.65 | 55.45 | 0.59 | 55.36 | 0.60 | 1.70 |
| School type | | | | | | | |
| Public | 53.68 | 0.32 | 53.49 | 0.32 | 53.49 | 0.33 | 11.81 |
| Private | 56.64 | 0.48 | 56.89 | 0.49 | 56.76 | 0.49 | 3.90 |
| Socioeconomic status quartiles | | | | | | | |
| 25% below | 47.70 | 0.30 | 47.88 | 0.32 | 47.77 | 0.31 | 1.83 |
| 25-50% | 50.28 | 0.27 | 50.26 | 0.29 | 50.22 | 0.29 | 5.01 |
| 50-75% | 54.40 | 0.28 | 54.45 | 0.29 | 54.41 | 0.29 | 3.50 |
| 75% above | 60.15 | 0.37 | 59.71 | 0.36 | 59.78 | 0.36 | 3.33 |

SOURCE: U.S. Department of Education, National Center for Education Statistics, Early Childhood Longitudinal Study, Kindergarten Class of 2010-11. 2010 Fall.

Results from Steps 2-5 provide the basis for the classification of observed predictors according to Table 1. Steps 2-3 show that the observed predictors are generally weakly related to the survey variables, and Step 5 finds that the response propensities are weakly related to the survey outcome. Referring to the first column in Table 1, weighting adjustment based on the observed predictors and the response propensities will not substantially affect nonresponse bias.

The resulting measure of deviation for the proxy variable is small with $d = 0.01$, $X$ is a weak proxy (with small $\hat{\rho}$), so the adjustment in the mean estimate $\hat{\mu}_y$ is small. This is some evidence against nonresponse bias, but this evidence is relatively weak since the correlation is weak.

**5.7. Step 7: Assess effects of nonresponse weighting adjustments on key survey estimates.**

We compare ECLS-K:2011 estimates of average reading IRT scores in Table 2 between unweighted and weighted estimates, and between estimates using coverage-adjusted base weights and estimates using nonresponse adjusted weights.

Sampling variance estimation has to account for design features such as clustering and survey weights. We use the Taylor series linearization approximation to obtain the standard errors with the PSU clustering effects and weights in the estimation (Binder, 1983). Table 2 also includes the design effects, calculated as the ratio of estimate variances under the complex design with PSUs and final weights, and the simple random sampling selection with the same sample sizes. For subgroup analyses, in addition to subgroups defined by race/ethnicity and school type, we add the estimates of subgroups defined by the quartiles of socioeconomic status. The estimates are significantly different across subgroups. API students tend to have better reading performances than those in different race/ethnicity groups. Students in private schools have higher scores than those in public schools. The reading scores are highly correlated with socioeconomic status.

The existing nonresponse adjustment of ECLS-K: 2011 uses weighting classes defined by the cross-tabulation of date of birth, race/ethnicity, and school type (Tourangeau et al., 2013). This shows that the nonresponse weighting factors will be equal for subjects that fall in each subgroup defined by race/ethnicity or school type. This is not the case for the subgroups defined by socioeconomic status. The unweighted, unadjusted, and adjusted estimates are not substantially different. The findings on nonresponse adjustment effects are consistent for overall and subgroup mean estimates. The complex sample design is considerably less efficient than simple random sampling. Because the weighted and unweighted estimates and standard errors are similar, the large design effects are mainly due to the PSU clustering effects, not due to weighting adjustments, which is confirmed by the large values when only accounting for clustering in the complex design.

## 5.8. Step 8: Compare the survey with external data using summary estimates of key survey variables.

Current NRBA by Westat compare estimates of selected items from the base year ECLS-K:2011 parent interviews and the parent interviews in the 2007 National Household Education Survey, subset to parents of kindergartners. The differences in the estimates between the two studies could be due to various sources of heterogeneity in data collection, such as time discrepancies, coverage, and sample design. These comparisons inform potential nonresponse bias but are not direct assessments. External data of high quality are crucial to validate the NRBA.

## 5.9. Step 9: Sensitivity analysis.

We set the sensitivity parameter $\phi$ as 0, 0.5 and 1, respectively, where $\phi = 0$ indicates MAR, $\phi = 1$ indicates an MNAR case where the missing-data mechanism only depends on the survey variable $Y$, and the mid-point $\phi = 0.5$ indicates MNAR where both survey and proxy variables affect the response propensity. We compare the different estimates of $\hat{\mu}_y$ to assess the deviation from MAR. Table 3 displays maximum likelihood estimates of the mean estimates under different missing data mechanisms, and the standard error estimates are based on large-sample approximations based on Equation (6). Alternatives approaches are Bayesian methods or multiple imputation (Andridge and Little, 2011).

**Table 3. NRBA for mean estimates (with standard errors in parenthesis) under different missingness mechanisms for different survey variables.**

|  | Correlation $\hat{\rho}$ | Deviation d | $\bar{y}_R$ | $\hat{\mu}_y$: MAR ($\phi = 0$) | $\hat{\mu}_y$: MNAR ($\phi = 0.5$) | $\hat{\mu}_y$: MNAR ($\phi = 1$) |
|---|---|---|---|---|---|---|
| Child |  |  |  |  |  |  |
| Reading IRT score | 0.36 | 0.01 | 54.07 | 54.10 (0.09) | 54.16 (0.10) | 54.30 (0.13) |
| Math IRT score | 0.44 | 0.01 | 35.56 | 35.61 (0.09) | 35.68 (0.09) | 35.83 (0.12) |
| BMI | 0.18 | -0.03 | 16.50 | 16.49 (0.02) | 16.44 (0.02) | 16.20 (0.04) |
| Parent |  |  |  |  |  |  |
| Impulsive/overactive | 0.18 | -0.02 | 2.04 | 2.04 (0.01) | 2.03 (0.01) | 2.00 (0.02) |
| Self-control | 0.13 | 0.01 | 2.89 | 2.89 (0.01) | 2.87 (0.01) | 2.75 (0.02) |
| Teacher |  |  |  |  |  |  |
| Externalizing | 0.28 | -0.01 | 1.60 | 1.60 (0.01) | 1.59 (0.01) | 1.54 (0.01) |
| Internalizing | 0.14 | 0.01 | 1.46 | 1.46 (0.004) | 1.46 (0.005) | 1.45 (0.01) |

SOURCE: U.S. Department of Education, National Center for Education Statistics, Early Childhood Longitudinal Study, Kindergarten Class of 2010-11. 2010 Fall.

**Table 4. NRBA of reading IRT scores for subgroups.**

|  | Correlation $\hat{\rho}_k$ | Deviation $d_k$ | $\bar{y}_{Rk}$ | $\hat{\mu}_{yk}$: MAR ($\phi = 0$) | $\hat{\mu}_{yk}$: MNAR ($\phi = 0.5$) | $\hat{\mu}_{yk}$: MNAR ($\phi = 1$) |
|---|---|---|---|---|---|---|
| Race/ethnicity |  |  |  |  |  |  |
| White (not Hisp) | 0.26 | 0.01 | 55.45 | 55.49 (0.13) | 55.61 (0.13) | 56.09 (0.23) |
| Black (not Hisp) | 0.27 | -0.001 | 52.25 | 52.26 (0.22) | 52.29 (0.23) | 52.38 (0.39) |
| Hispanic | 0.27 | -0.05 | 50.32 | 50.19 (0.16) | 49.83 (0.16) | 48.54 (0.27) |
| API (not Hisp) | 0.40 | -0.01 | 58.17 | 58.07 (0.41) | 57.93 (0.43) | 57.59 (0.60) |
| Other | 0.31 | 0.01 | 56.25 | 56.30 (0.48) | 56.40 (0.49) | 56.74 (0.62) |
| School type |  |  |  |  |  |  |
| Public | 0.35 | 0.01 | 53.68 | 53.72 (0.10) | 53.80 (0.10) | 54.02 (0.15) |
| Private | 0.36 | -0.002 | 56.64 | 56.63 (0.25) | 56.62 (0.25) | 56.58 (0.33) |

SOURCE: U.S. Department of Education, National Center for Education Statistics, Early Childhood Longitudinal Study, Kindergarten Class of 2010-11. 2010 Fall.

For the reading IRT scores, the mean estimates in Table 3 slightly increase as $\phi$ increases, and the standard error of the case with $\phi = 1$ is larger than the remaining two values. The estimates are similar to each other and not sensitive to different missingness mechanisms, providing some evidence that the potential nonresponse bias is small. Again, the evidence is weak because the proxy variable is not strongly related to the outcome.

Table 3 also presents the NRBA results for other outcomes of interest. The correlations between the outcome and the proxy variables are generally low, ranging from 0.13 to 0.44, showing modest evidence for the NRBA. The sensitivity analyses find that the mean estimates for BMI, scores for self-control and externalizing problems are significantly different under different missing data mechanisms, where the nonrespondents have substantially lower average scores than respondents. This indicates the potential for nonresponse bias for these outcomes, but the evidence is modest.

We estimate average reading IRT scores across subgroups defined by race/ethnicity and school type in Table 4. Similar to those overall estimates, subgroup estimates are not substantially different under different missing data mechanisms. Even with subtle differences, we observe different adjustment effects across subgroups as the missing data mechanisms change in the sensitivity analysis.

**5.10. Step 10: Item nonresponse bias analysis.**

Continuing the example with the reading IRT score, the outcome has 50 more null values than the number of unit nonresponses and 120 values of -9 indicating item nonresponse. Since the proportion of nonresponse mismatch is very small, we treat the 15,670 observed cases of the outcome variable as respondents. However, item nonresponse could also lead to substantial bias. Future extensions of this work will perform MI and assess the effects on inferences.

**6. Discussion**

We present a ten-step exemplar of the NRBA for cross-sectional studies. Our NRBA assesses the pattern of missing data, fits regressions of key survey outcomes and indicators of nonresponse on variables observed for both respondents and nonrespondents, compares estimates with and without nonresponse weighting adjustments, and implements sensitivity analyses based on proxy pattern-mixture models to assess the impact of deviations from MAR missingness. All analyses can be carried out straightforwardly with standard statistical software. We provide our example R codes in the Appendix.

Overall, we do not find substantial evidence of nonresponse bias in the ECLS-K:2011 study, though modest differences present for several estimates, perhaps reflecting the high response rates at baseline. However, lack of evidence of bias in the NRBA does not necessarily mean lack of bias; the key to a strong NRBA is the existence of a rich set of auxiliary variables that are highly predictive of the survey variables. The strength of the evidence is generally weak in this application because the observed predictors are not strongly related to the survey outcomes. The auxiliary variables are collected from the frame, available for both respondents and nonrespondents, but a time lag exists between the frame (around 2007) and the actual data collection (around 2010), leading to weak correlations with the survey variables. The ECLS-K:2011 dataset has the children's zip codes and geographic information that can be linked to the census data for the neighborhood characteristics. Obtaining auxiliary information via geospatial linking will be future work.

As regards future work, data integration with multiple sources greatly enhances many ongoing survey research activities (NASEM, 2017). The NRBA requires information that is available for both respondents and nonrespondents. Combining multiple data sources and record linkage can provide auxiliary information for nonresponse adjustment and benchmark information for external validation. The proxy pattern-mixture models assume that the response mechanism depends on an additive effect of the survey variable and the proxy variable, and the assumption cannot be verified without external data. The mean estimates with large sample sizes are robust against the normality misspecification (Andridge and Thompson, 2015), but the effect on other estimands with small samples is unclear and needs future work. Linking ECLS-K:2011 studies to other data could have great potential for the NRBA. Second, the NRBA of analytic inferences could have different findings from that of descriptive summaries.

We focus here on mean estimates for the population and population subgroups. The NRBA in regression models is important, and many of the steps outlined above can also be applied in the regression setting. Extensions of the proxy pattern-mixture approach to sensitivity analysis for regression are discussed in West et al. (2021). In the regression setting, the key to a strong analysis is the availability of strong auxiliary variables that are not predictors in the regression model of interest. Third, we mainly demonstrate the assessment and adjustment of estimates for unit nonresponse, which is the main concern in ECLS-K; assessment of item nonresponse may be important in surveys where the level of item nonresponse is greater. Si et al. (2021) has demonstrated the applicability of MI in handling various challenges on item nonresponse with massive datasets. MI has been used to simultaneously handle unit nonresponse and item nonresponse (Si, Reiter, and Hillygus, 2015; 2016). Combining weighting adjustment and imputation into one systematic process would be helpful for practical survey operation. Future work will also develop an NRBA exemplar for longitudinal studies.

Appendix: Computational R codes for NRBA steps based on the ECLS-K data.

##Step 1: Analyze missing-data patterns

## Steps 2-4: model selected key survey variables

#We identified 7 key survey variables and use the reading IRT scores (X1RSCALK5) as an illustration.

#use tree-based mdodesl for variable selection
```
Y.t = ctree(X1RSCALK5 ~  C2BTW0 + X12RACETH + X_CHSEX + X_DOBYY + X1KSCTYP + X1REGION+
X1LOCALE + FTE + NUMSTU + factor(LOGR) + factor(HIGR) + P_KG + P_AM + P_AS + P_HS +
P_BL + STR,data = ecls_stu)
  Y.m0 = lm(X1RSCALK5 ~ 1, data = ecls_stu)
  #add two-way interactions selected from the tree model into regression
  Y.m1 = lm(X1RSCALK5 ~ C2BTW0 + X12RACETH + X_CHSEX + X_DOBYY + X1KSCTYP + X1REGION +
    X1LOCALE + FTE + NUMSTU + LOGR + HIGR +
    P_KG + P_AM + P_AS + P_HS + P_BL + STR + X12RACETH * X1REGION +
    X1LOCALE * FTE + P_HS * FTE + LOGR * X12RACETH + P_BL * X12RACETH +
    P_AS * X12RACETH + NUMSTU * HIGR + X12RACETH * X1KSCTYP +
    P_HS * X12RACETH + X1LOCALE * X1KSCTYP + NUMSTU * X1KSCTYP + P_AS * X1LOCALE +
X_DOBYY * X1REGION + X_DOBYY * P_AS + X_DOBYY * P_HS,
  data = ecls_stu)

  #step-wise variable selection
  Y.m = stepAIC(Y.m0,scope = list(lower = Y.m0, upper = Y.m1), direction = "forward",
trace = FALSE)

  #save predicted values as a proxy variable
  ecls_stu = ecls_stu %>% mutate(X1RSCALK5.p = predict(Y.m, ecls_stu))
```

## Step 5: model response propensities

```r
#tree model for variable selection
I.st.tr1 = ctree(C1STAT ~ C2BTW0 + X12RACETH + X_CHSEX + X_DOBYY + X1KSCTYP + X1REGION
+ X1LOCALE + FTE + NUMSTU + factor(LOGR) + factor(HIGR) +P_KG + P_AM + P_AS + P_HS +
P_BL + STR, data = stu_nr)
I.st1.n = glm(C1STAT ~ 1, data = stu_nr, family = "binomial")
I.st1.f = glm(C1STAT ~ C2BTW0 + X12RACETH + X_CHSEX + X_DOBYY + X1KSCTYP + X1REGION +
X1LOCALE + FTE + NUMSTU + LOGR + HIGR +
P_KG + P_AM + P_AS + P_HS + P_BL + STR + X12RACETH * X1REGION + X1LOCALE * FTE + P_HS
* FTE + LOGR * X12RACETH +P_BL * X12RACETH + P_AS * X12RACETH + NUMSTU * HIGR,
data = stu_nr,family = "binomial")

I.st1 = stepAIC(I.st1.n, scope = list(lower = I.st1.n, upper =  ~ (
C2BTW0 + X12RACETH + X_CHSEX + X_DOBYY + X1KSCTYP + X1REGION + X1LOCALE + FTE + NUMSTU
+ LOGR + HIGR + P_KG + P_AM + P_AS + P_HS + P_BL + STR + X12RACETH * X1REGION +
X1LOCALE * FTE + P_HS * FTE + LOGR * X12RACETH + P_BL * X12RACETH + P_AS * X12RACETH +
NUMSTU * HIGR)), direction = "forward", trace = FALSE)

#Predicted response propensities
r.s = predict(I.st1, type = "response")

#AUC
auc(roc(stu_nr$C1STAT ~ r.s, plot = FALSE, print.auc = TRUE))

#save predicted response propensities and create a stratification indicator based on
quintiles
ecls_stu = ecls_stu %>% mutate(I.st1.p = predict(I.st1, ecls_stu, type = "response"),
I.st1.p.s = cut(I.st1.p,breaks = quantile(I.st1.p, probs = seq(0, 1, by = 0.2)),
include.lowest = TRUE))
```

## Step 6: assess the measure of deviation for the proxy variable

```r
 (mean(ecls_stu$X1RSCALK5.p) - mean(ecls_stu$X1RSCALK5.p[ecls_stu$C1STAT ==
  1]))/sd(ecls_stu$X1RSCALK5.p[ecls_stu$C1STAT==1])
```

## Step 7: Assess nonresponse weighting adjustment effects

```
#define survey objects
now_svy = svydesign(ids = ~ PSUID, data = ecls_stu[!is.na(ecls_stu$W1C0), ])
c2btw0_svy = svydesign(ids = ~ PSUID, weights =  ~ C2BTW0, data = ecls_stu[!is.na(ecls_stu$W1C0), ])
w1c0_svy = svydesign(ids = ~ PSUID, weights =  ~ W1C0, data = ecls_stu[!is.na(ecls_stu$W1C0), ])

#overall mean
svymean(~ X1RSCALK5, now_svy, deff = T, na.rm = TRUE)
svymean(~ X1RSCALK5, c2btw0_svy, deff = T, na.rm = TRUE)
svymean(~ X1RSCALK5, w1c0_svy, deff = T, na.rm = TRUE)

#subgroup analysis: by race/ethnicity
svymean(~ X1RSCALK5, subset(now_svy, X12RACETH == 1), deff = T, na.rm = TRUE)
svymean(~ X1RSCALK5,subset(c2btw0_svy, X12RACETH == 1),deff = T,na.rm = TRUE)
svymean(~ X1RSCALK5, subset(w1c0_svy, X12RACETH == 1), deff = T, na.rm = TRUE)
```

## Step 9: sensitivity analysis

Function to get the MLE under proxy pattern mixture models

```
mle <- function(x, y, lambda){
  # Indicator for missingness
  m <- ifelse(is.na(y), 1, 0)
  # Number of respondents, nonrespondents
  n <- length(x); r <- sum(1-m)
  # Calculate means, sums of squares # Respondent data
  xBar_0 <- mean(x[m==0]); yBar_0 <- mean(y[m==0]);
sxx_0 <- sum((x[m==0] - xBar_0)^2)/r; syy_0 <- sum((y[m==0] - yBar_0)^2)/r;
  sxy_0 <- sum((x[m==0] - xBar_0) * (y[m==0] - yBar_0))/r;
  byx.x_0 <- sxy_0 / sxx_0; by0.x_0 <- yBar_0 - byx.x_0*xBar_0;
  syy.x_0 <- syy_0 - (sxy_0)^2 / sxx_0; bxy.y_0 <- sxy_0 / syy_0;
  sxx.y_0 <- sxx_0 - (sxy_0)^2 / syy_0; rhoHat_0 <- sxy_0/sqrt(sxx_0*syy_0);
  # Nonrespondent data
```

```r
xBar_1 <- mean(x[m==1]); sxx_1 <- sum((x[m==1] - xBar_1)^2)/(n-r)

# Value of g_lambda
if (lambda==Inf) {
  g.lambda <- 1/bxy.y_0
} else {
  g.lambda <- sqrt(syy_0/sxx_0)*(lambda + rhoHat_0)/(lambda*rhoHat_0 + 1)
  # Note that this reduces to byx.x_0=sxy_0/sxx_0 when lambda=0   }

# Variance of g_lambda
if (lambda==Inf) {
  num <- (sxx_0*syy_0 - sxy_0^2)*syy_0^2
  denom <- r*sxy_0^4;  g.lambda.var <- num/denom
} else {
  a <- sxx_0^2*syy_0^2*(1-lambda^2+lambda^4)
  b <- 2*sxx_0*syy_0*sxy_0*lambda*(3*lambda*sxy_0 + sqrt(sxx_0*syy_0)*(1+lambda^2))
  c <- lambda*sxy_0^3*(lambda*sxy_0 + 2*sqrt(sxx_0*syy_0)*(1+lambda^2))
  num <- (sxx_0*syy_0 - sxy_0^2)*(a + b + c)
  denom <- r*sxx_0^2*(sqrt(sxx_0*syy_0) + lambda*sxy_0)^4
  g.lambda.var <- num/denom}
# MLE of E(x)
muX <- mean(x);
# MLE of Var(x)
sigmaXX <- (r/n)*sxx_0 + ((n-r)/n)*sxx_1 + (r/n)*((n-r)/n)*(xBar_0 - xBar_1)^2;

# MLE of E(y)
muY <- yBar_0 + g.lambda*(muX - xBar_0);
# MLE of Var(y)
sigmaYY <- syy_0 + (g.lambda^2)*(sigmaXX - sxx_0);
# MLE of Cov(x,y)
sigmaXY <- sxy_0 + g.lambda*(sigmaXX - sxx_0);
# Variance of MLE of E(y)
one <- sigmaYY/n; two <- g.lambda.var*(muX - xBar_0)^2;
```

```
  three <- ((n-r)/(r*n)) *(syy_0 - 2*g.lambda*sxy_0 + g.lambda^2*sxx_0);
  muYvar <- one + two + three
  return(list(muY=muY, muYvar=muYvar))}

#Correlation between predicted values (proxy) and outcomes
cor(ecls_stu$X1RSCALK5.p,ecls_stu$X1RSCALK5, use = "complete.obs")

#MAR ($\phi = 0$): the mean value is
cat(round(mle(ecls_stu$X1RSCALK5.p, ecls_stu$X1RSCALK5, 0)$muY,2), " with se=",
round(sqrt(mle(ecls_stu$X1RSCALK5.p, ecls_stu$X1RSCALK5, 0)$muYvar),2))  # lambda = 0

#MNAR ($\phi = 0.5$): the mean value is
cat(round(mle(ecls_stu$X1RSCALK5.p, ecls_stu$X1RSCALK5, 1)$muY,2), " with se=",
round(sqrt(mle(ecls_stu$X1RSCALK5.p, ecls_stu$X1RSCALK5, 1)$muYvar),2))   # lambda = 1

#MNAR ($\phi = 1$): the mean value is
cat(round(mle(ecls_stu$X1RSCALK5.p, ecls_stu$X1RSCALK5, Inf)$muY,2), " with se=",
round(sqrt(mle(ecls_stu$X1RSCALK5.p, ecls_stu$X1RSCALK5, Inf)$muYvar),2))
```